\documentclass{WCCM-ECCOMAS_2026_Full_Paper}

\usepackage{graphicx}
\usepackage{amsmath}
\usepackage{booktabs}
\usepackage{float}
\usepackage[colorlinks=true,linkcolor=blue,urlcolor=blue,citecolor=blue]{hyperref}
\usepackage{subcaption}

\title{Online Learning of a Spectral Viscosity Closure for Large Eddy Simulation Using ensemble Kalman Inversion}

\author{Katerina Kostova$^{1}$, Yifei Guan$^{1\ast}$}
\shortauthors{Kostova and Guan}

\address{%
$^{1}$ \textit{Department of Mechanical Engineering, Union College, Schenectady, New York, USA}\\
$^{*}$Corresponding author: \textit{guany@union.edu}}


\summary{Large Eddy Simulation (LES) resolves only the large, energy-containing structures of turbulent flows while modeling the effects of unresolved subgrid scales. As a result, the accuracy of LES strongly depends on the subgrid-scale closure. This work investigates the online calibration of a spectral viscosity closure for forced two-dimensional turbulence using ensemble Kalman Inversion (EKI). The closure consists of a wavenumber-dependent viscosity function, which is parameterized by a 16-dimensional vector. An ensemble of potential parameter vectors is passed through coarse resolution LES simulations, and the resulting kinetic energy spectra are compared with a DNS reference spectrum up to the LES-resolved wavenumber range. Then, ensemble statistics iteratively update the spectral viscosity parameters in order to reduce the model-data discrepancy. The convergence of the spectral mismatch is used to assess the accuracy of the calibrated closure by comparing LES with the filtered DNS data in spectral space. The results demonstrate the potential of EKI-based online calibration as a derivative-free framework for constructing data-driven turbulence closures with multiple parameters.}

\begin{document}

\maketitle

\section{Introduction}

Turbulent flows contain a wide range of interacting spatial and temporal scales, making their numerical simulation computationally demanding. Direct Numerical Simulation (DNS) resolves the turbulent scales that are relevant to the spatial dynamics and can therefore provide high-fidelity representations of turbulent flows. However, the computational cost of DNS increases rapidly with Reynolds number, which limits its applicability to many practical engineering and geophysical problems. Large Eddy Simulation (LES) provides a more efficient and less computationally expensive alternative by explicitly resolving the larger, energy-containing structures while modeling the influence of the unresolved smaller scales, commonly referred to as subgrid scales (SGS). Consequently, the choice of SGS closure plays an important role in determining the accuracy and physical fidelity of an LES solution.

One of the most established approaches to SGS modeling is the eddy-viscosity hypothesis. The classical Smagorinsky model introduces a positive eddy viscosity which is related to the local strain rate. The model provides a robust mechanism for transferring kinetic energy from resolved to unresolved scales. Although its dissipative character contributes to numerical stability, it cannot effectively represent more complex energy transfer between scales, such as backscatter from unresolved to resolved scales, and, therefore, may lead to excessive dissipation of resolved structures. Backscatter-capable closures attempt to address this limitation by permitting bidirectional energy transfer, although doing so may introduce additional challenges for numerical stability \cite{GuanCNN2022,GuanEKI2026}.

An alternative approach is spectral viscosity, in which additional dissipation is defined directly in Fourier space as a function of the wavenumber. This allows the closure to act at different strengths across spatial scales, with weaker influence on larger resolved structures and stronger dissipation toward the high wavenumber portion of the resolved spectrum. Spectral vanishing viscosity was originally introduced in the context of nonlinear conservation laws and was subsequently extended to LES of turbulent flows \cite{Tadmor1989,Yu1999,Karamanos2000,Pasquetti2006}. The resulting wavenumber-dependent dissipation provides a mechanism for stabilizing high-wavenumber modes of the smaller resolvable scales while preserving the larger resolved structures~\cite{Matharu2022}.

Nonetheless, the benefit of flexibility allowed by a wavenumber-dependent spectral viscosity creates an additional challenge: determining an appropriate viscosity profile for the turbulent flow under consideration. Rather than prescribing this profile a priori, data-driven calibration provides a means of inferring closure parameters from high-fidelity reference data. Ensemble Kalman Inversion (EKI) offers a derivative-free approach to this inverse problem by evolving an ensemble of candidate parameter vectors according to discrepancies between model predictions and observations \cite{iglesias2013ensemble,Cleary2021}. Previous work has confirmed that EKI can be used for online calibration of physics-based eddy-viscosity and backscatter closures in geophysical turbulence \cite{GuanEKI2026}. This approach retains a closure that is physically interpretable, while using high-fidelity data to construct its parameters.

This paper presents an online EKI calibration of a parameterized spectral viscosity closure for forced two-dimensional turbulence. The case explored features a spectral viscosity, represented by a 16-dimensional parameter vector, which controls its dissipation strength. An ensemble of candidate parameter vectors is propagated through coarse-resolution LES simulations, and the resulting kinetic energy spectra are compared with a DNS reference spectrum, bounded by the wavenumber range used for the LES solution. The ensemble is iteratively updated using data discrepancies of the model as well as ensemble covariance information. The resulting closure is evaluated through the convergence of the calibration procedure. Furthermore, comparisons are made between the calibrated LES and the reference solution to confirm the accuracy of the results.

\section{Mathematical formulation}

\subsection{Forced two-dimensional turbulence}

The testbed for this numerical investigation is built upon the forced two-dimensional incompressible turbulence on a periodic domain. In vorticity form, its governing equation is the following:
\begin{equation}
\frac{\partial \omega}{\partial t}
+
\mathbf{u}\cdot\nabla\omega
=
\nu\nabla^2\omega
-
\alpha\omega
+
f,
\label{eq:vorticity}
\end{equation}
where $\omega$ represents the scalar vorticity, $\mathbf{u}=(u,v)$ is the incompressible velocity field, $\nu$ is the molecular viscosity, $\alpha$ is a linear-drag coefficient, and $f$ is an external forcing term. The advection term describes the transport of vorticity by the evolving velocity field. Meanwhile, viscous dissipation and large-scale damping simultaneously affect this evolution by smoothing out vorticity variations and removing energy from the flow. Finally, the external forcing continuously injects energy into the system to sustain its turbulent characteristic. The divergence-free velocity field is recovered through the streamfunction $\psi$,
\begin{equation}
\nabla^2\psi=-\omega,
\label{eq:streamfunction}
\end{equation}
with
\begin{equation}
\mathbf{u}
=
\left(
\frac{\partial\psi}{\partial y},
-\frac{\partial\psi}{\partial x}
\right).
\label{eq:velocity}
\end{equation}

The Fourier spectral discretization of the governing equations is achieved using periodic boundary conditions, while the individual Fourier modes are determined by the wavevector, defined as 
\begin{equation}
\mathbf{k}=(k_x,k_y),
\end{equation}
with radial wavenumber
\begin{equation}
k=|\mathbf{k}|=\sqrt{k_x^2+k_y^2}.
\label{eq:wavenumber}
\end{equation}
The smaller the $k$-values, the larger the spatial structures. The opposite also holds true, as increasing $k$ corresponds to progressively smaller scales.

\subsection{Spectral viscosity parameterization}

Coarse-resolution LES struggles to explicitly resolve the smallest turbulent scales that have been captured by the DNS. Therefore, an alternative is to model those scales using a spectral viscosity closure. The spectral viscosity can be denoted by
\begin{equation}
\nu_t(k;\boldsymbol{\theta}),
\end{equation}
where
\begin{equation}
\boldsymbol{\theta}
=
(\theta_1,\theta_2,\ldots,\theta_{p})^T
\in \mathbb{R}^{p}
\label{eq:theta}
\end{equation}
is a $p$-dimensional parameter vector. Here, we choose $p=16$ as the number of control points. Altogether, the coefficients shape the viscosity profile over the resolved wavenumber range.

In the numerical implementation, the 16 parameters define the spectral viscosity profile without prescribing where dissipation is concentrated. EKI therefore determines both the magnitude and scale distribution of the added dissipation from the FDNS statistics. In Fourier space, the spectral viscosity supplements the molecular viscosity. Since the Laplacian operator is represented by a multiplication by $-k^2$ in Fourier space, the overall contributing dissipation to each vorticity mode can be written as

\begin{equation}
-\left[\nu + \nu_t(k;\boldsymbol{\theta})\right]k^2\hat{\omega}.
\end{equation}

Here, $\hat{\omega}$ denotes the Fourier coefficient of the vorticity field, $\nu$ is the molecular viscosity, and $\nu_t(k;\boldsymbol{\theta})$ is the parameterized spectral viscosity. This term applies the learned scale-dependent dissipation independently to each Fourier mode/wavenumber.

\begin{figure}[H]
\centering
\includegraphics[width=0.78\textwidth]{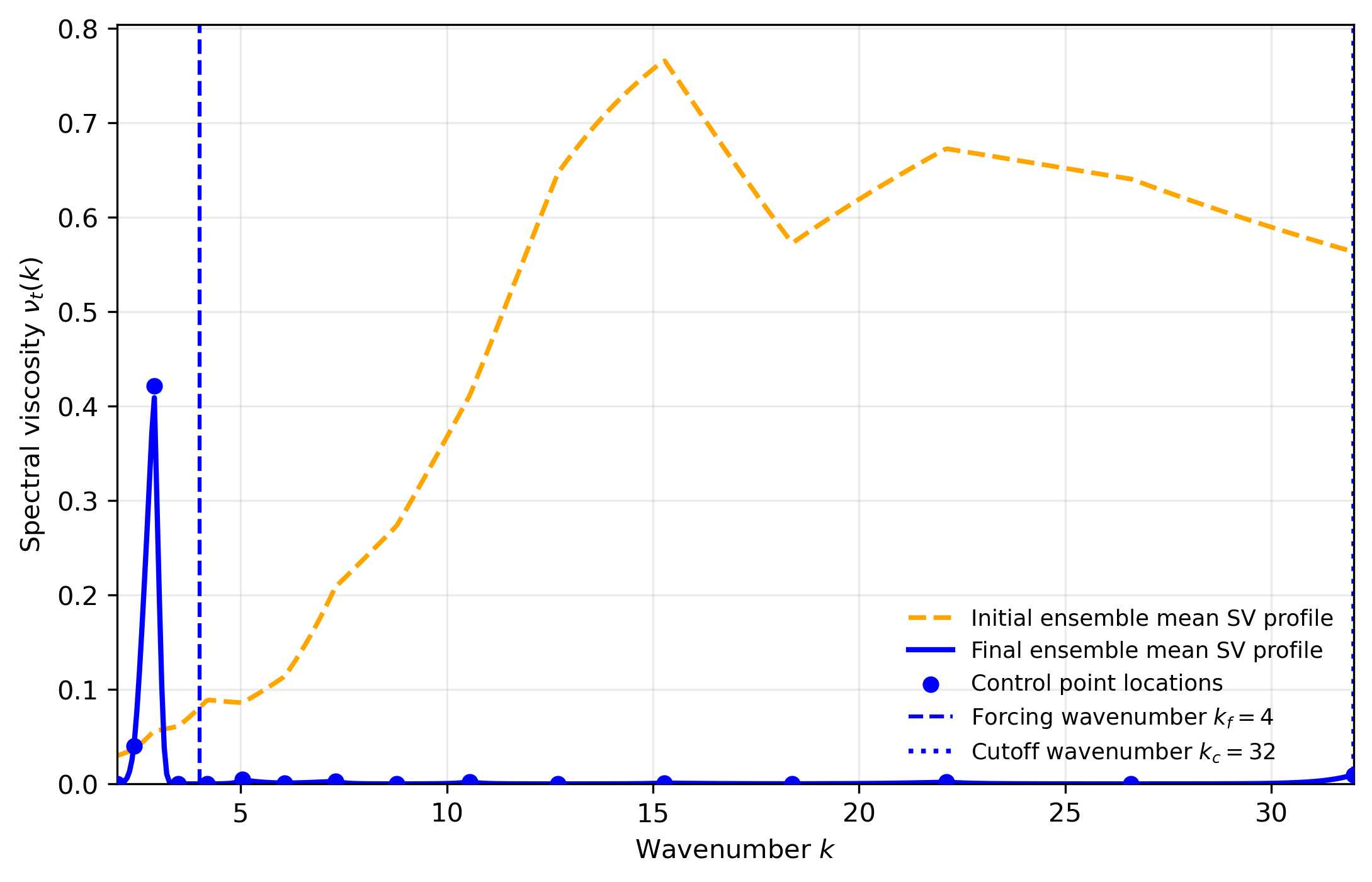}
\caption{Initial and final ensemble mean spectral viscosity profiles $\nu_t(k;\boldsymbol{\theta})$. The 16 control point locations parameterize the closure over the LES resolved wavenumber range. The forcing wavenumber $k_f=4$ and LES cutoff wavenumber $k_c=32$ are indicated for reference.}
\label{fig:sv}
\end{figure}

Furthermore, the mapping between the parameter vector and the resulting LES statistics is represented by the forward model
\begin{equation}
G(\boldsymbol{\theta}),
\label{eq:forward}
\end{equation}
where $G$ denotes the LES solver with the spectral viscosity model governed by parameters $\boldsymbol{\theta}$ (e.g., spectral viscosity), followed by computation of the kinetic energy spectrum used for calibration. Figure~\ref{fig:sv} compares the initial and final ensemble-mean spectral viscosity profiles. Instead of reproducing a conventional spectral vanishing-viscosity shape, the learned profile concentrates its strongest dissipation near \(k\approx 3\), close to the forcing wavenumber and inverse cascade range. This distribution is determined entirely by the EKI calibration against the FDNS statistics.

\section{Ensemble Kalman Inversion calibration}

\subsection{Inverse problem and ensemble representation}

The objective of the calibration step is to infer the spectral viscosity parameters that minimize the discrepancy between LES statistics and the high-resolution reference DNS data. The inverse problem is written as
\begin{equation}
G(\boldsymbol{\theta})
\approx
\mathbf{y}_{\mathrm{FDNS}},
\label{eq:inverse}
\end{equation}
where $\mathbf{y}_{\mathrm{FDNS}}$ denotes the DNS kinetic energy spectrum restricted to the LES wavenumber range.

The ensemble used in EKI is composed as follows:
\begin{equation}
\left\{
\boldsymbol{\theta}_n^{(j)}
\right\}_{j=1}^{J},
\end{equation}
where $j$ refers to the ensemble member and $n$ denotes an EKI iteration. Overall, each of these ensemble members represents a complete 16-dimensional candidate parameter vector. The ensemble is initialized from a normal distribution
\begin{equation}
\boldsymbol{\theta}_0^{(j)}
\sim
\mathcal{N}(\mathbf{m}_0,\Sigma_0),
\qquad
j=1,\ldots,J,
\label{eq:prior}
\end{equation}
where $\mathbf{m}_0$ and $\Sigma_0$ denote the prior mean and covariance.

At each EKI iteration, every ensemble member is individually passed through the LES forward model,
\begin{equation}
\mathbf{y}_n^{(j)}
=
G\left(\boldsymbol{\theta}_n^{(j)}\right).
\label{eq:forward_member}
\end{equation}

\subsection{Ensemble statistics}

The parameter and model-prediction means are
\begin{equation}
\bar{\boldsymbol{\theta}}_n
=
\frac{1}{J}
\sum_{j=1}^{J}
\boldsymbol{\theta}_n^{(j)},
\label{eq:theta_mean}
\end{equation}
and
\begin{equation}
\bar{\mathbf{G}}_n
=
\frac{1}{J}
\sum_{j=1}^{J}
G\left(\boldsymbol{\theta}_n^{(j)}\right).
\label{eq:G_mean}
\end{equation}

Additionally, the parameter covariance, which describes the spread and correlations of the parameter ensemble, can be evaluated as
\begin{equation}
C_{\theta,n}
=
\frac{1}{J}
\sum_{j=1}^{J}
\left(
\boldsymbol{\theta}_n^{(j)}
-
\bar{\boldsymbol{\theta}}_n
\right)
\left(
\boldsymbol{\theta}_n^{(j)}
-
\bar{\boldsymbol{\theta}}_n
\right)^T,
\label{eq:Ctheta}
\end{equation}
while the parameter-model cross covariance is
\begin{equation}
C_{\theta G,n}
=
\frac{1}{J}
\sum_{j=1}^{J}
\left(
\boldsymbol{\theta}_n^{(j)}
-
\bar{\boldsymbol{\theta}}_n
\right)
\left(
G(\boldsymbol{\theta}_n^{(j)})
-
\bar{\mathbf{G}}_n
\right)^T.
\label{eq:CthetaG}
\end{equation}

The cross covariance describes the  relationship between the changes in a particular closure parameter and changes in the predicted spectrum. This allows this method to be derivative-free because the parameter ensemble is updated without explicitly differentiating the LES solver with respect to the 16 closure parameters.

\subsection{Adaptive EKI update}

Furthermore, an adaptive EKI timestep is introduced to control the update magnitude according to how strongly an LES spectrum of an ensemble member differs from the FDNS target. The timestep uses a weighted measure of the discrepancy between the model and the empirical data. For ensemble member $j$,
\begin{equation}
r_n^{(j)}
=
\sqrt{
\left(
G(\boldsymbol{\theta}_n^{(j)})
-
\mathbf{y}_{\mathrm{FDNS}}
\right)^T
\Gamma_y^{-1}
\left(
G(\boldsymbol{\theta}_n^{(j)})
-
\mathbf{y}_{\mathrm{FDNS}}
\right)
},
\label{eq:misfit_member}
\end{equation}
where $\Gamma_y$ represents the observation-error covariance.

The EKI timestep is selected according to the largest mismatch present in the ensemble.
\begin{equation}
\Delta t_n
=
\operatorname{clip}
\left(
\frac{1}{\max_j r_n^{(j)}},
\Delta t_{\min},
\Delta t_{\max}
\right),
\label{eq:adaptive_dt}
\end{equation}
with
\begin{equation}
\Delta t_{\min}=10^{-6},
\qquad
\Delta t_{\max}=10.
\label{eq:dt_bounds}
\end{equation}

The update first computes an intermediate parameter vector,
\begin{equation}
\boldsymbol{\theta}_*^{(j)}
=
\left(
I+
\Delta t_n C_{\theta,n}\Sigma_0^{-1}
\right)^{-1}
\left[
\boldsymbol{\theta}_n^{(j)}
-
\Delta t_n
C_{\theta G,n}
\Gamma_y^{-1}
\left(
G(\boldsymbol{\theta}_n^{(j)})
-
\mathbf{y}_{\mathrm{FDNS}}
\right)
\right],
\label{eq:theta_star}
\end{equation}
to which a covariance-scaled random perturbation, called a stochastic EKI perturbation is added:
\begin{equation}
\boldsymbol{\theta}_{n+1}^{(j)}
=
\boldsymbol{\theta}_*^{(j)}
+
\sqrt{2\Delta t_n}\,
L_C\boldsymbol{\xi}_n^{(j)}.
\label{eq:stochastic_update}
\end{equation}
Here, $L_C$ is the Cholesky factor of the current ensemble parameter covariance, satisfying approximately
\begin{equation}
C_{\theta,n}
=
L_C L_C^T,
\end{equation}
and
\begin{equation}
\boldsymbol{\xi}_n^{(j)}
\sim
\mathcal{N}(0,I)
\end{equation}
is a standard Gaussian random vector. Following the deterministic correction, the stochastic perturbation is added to maintain ensemble spread rather than immediately collapsing to a single point. Meanwhile, the data mismatch term drives the ensemble toward parameter values producing improved agreement with the reference spectrum.

\section{Numerical Setup and Calibration Configuration}

\subsection{DNS and LES configuration}

 The reference and coarse-resolution simulations solve the same forced two-dimensional Navier-Stokes system. However, DNS does this by resolving a substantially wider range of turbulent scales, while the LES resolves the system at a far lower spatial resolution, which is why it requires a spectral viscosity closure. The DNS uses a $1024^2$ computational grid, while the LES is performed on a $64^2$ grid. Both simulations are defined on the periodic domain
\begin{equation}
L=2\pi,
\end{equation}
with molecular viscosity and Reynolds number
\begin{equation}
\nu=5\times10^{-5}, Re=1/\nu=20,000,
\end{equation}
linear drag coefficient
\begin{equation}
\alpha=0.1,
\end{equation}
and Kolmogorov forcing at forcing wavenumber
\begin{equation}
k_f=4.
\end{equation}

All variables and parameters are nondimensional. The DNS is performed without the learned spectral viscosity contribution, while the LES includes the parameterized closure. The governing equations are discretized using a Fourier pseudo-spectral method. The nonlinear advection term is advanced using a second-order Adams-Bashforth scheme, while the viscous term uses a Crank-Nicolson scheme. The spectral viscosity is also treated using the Crank-Nicolson scheme, as in Fourier space, the spectral viscosity commutes with the Laplacian operator.

\subsection{Reference spectrum and resolution matching}

The two-dimensional Fourier modes are grouped according to their radial wavenumber magnitude $k$ from Eq.~\eqref{eq:wavenumber}. Modes with approximately equal $k$ belong to the same wavenumber shell. Then, the kinetic energy contributions of the Fourier modes in each shell are summed to construct the one-dimensional kinetic energy spectrum $E(k)$.

For calibration, both spectra are restricted to the circular range $k\le N_{\mathrm{LES}}/2=32$; Cartesian corner modes with $k>32$ are not included in the EKI observation vector.

\subsection{Calibration target}

Because the $1024^2$ DNS represents substantially smaller spatial scales than the $64^2$ LES, the complete DNS spectrum is not used directly as the calibration target. Instead, the high-resolution DNS spectrum is truncated to
\begin{equation}
k=0,1,\ldots,\frac{N_{\mathrm{LES}}}{2}.
\end{equation}
For $N_{\mathrm{LES}}=64$, this retains
\begin{equation}
k=0,1,\ldots,32,
\end{equation}
or 33 shell values including $k=0$. This restricted spectrum is denoted here as the filtered DNS (FDNS) target.

The LES spectrum is averaged over multiple physical-time samples to reduce sensitivity to instantaneous turbulent fluctuations. The resulting spectrum forms the forward-model output used by EKI.

\subsection{Calibration configuration}

Each candidate closure is represented by the 16-dimensional parameter vector in Eq.~\eqref{eq:theta}. The initial parameter vectors are drawn from the prior distribution in Eq.~\eqref{eq:prior}. The implementation uses a zero prior mean and identity prior covariance,
\begin{equation}
\mathbf{m}_0=\mathbf{0},
\qquad
\Sigma_0=I.
\end{equation}

The EKI calibration uses an ensemble of $J=32$ members and is run over 16 update iterations. The 16 spectral viscosity control points are logarithmically distributed over $2 \leq k \leq 32$, using linear interpolation in $\log(k)$ between control locations, and the resulting
spectral viscosity is constrained to be nonnegative. The LES timestep is $\Delta t=5\times10^{-5}$ without explicit spectral de-aliasing. Each LES realization first undergoes re-equilibration for $10^5$ timesteps before any statistics are collected. The kinetic energy spectrum is then averaged over another $10^5$ timesteps and sampled every 200 timesteps, resulting in a total of 500 spectral samples for each averaged prediction.

\section{Results}

\subsection{EKI convergence}

\begin{figure}[H]
    \centering

    \begin{subfigure}[t]{0.38\textwidth}
        \centering
        \includegraphics[width=\linewidth]{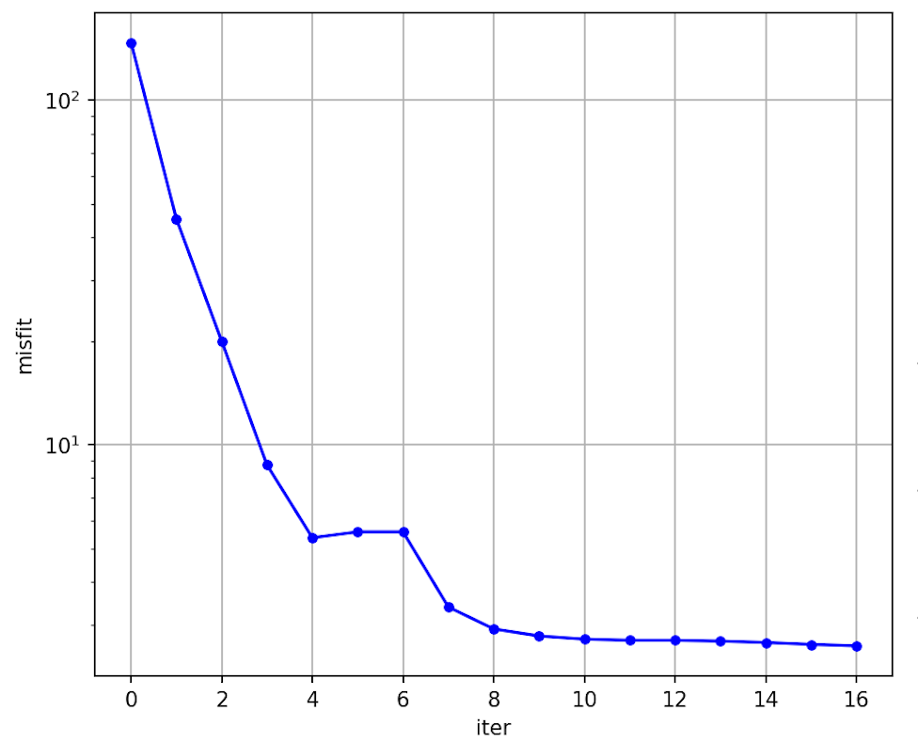}
        \caption{Spectral misfit.}
        \label{fig:eki_misfit}
    \end{subfigure}
    \hfill
    \begin{subfigure}[t]{0.48\textwidth}
        \centering
        \includegraphics[width=\linewidth]{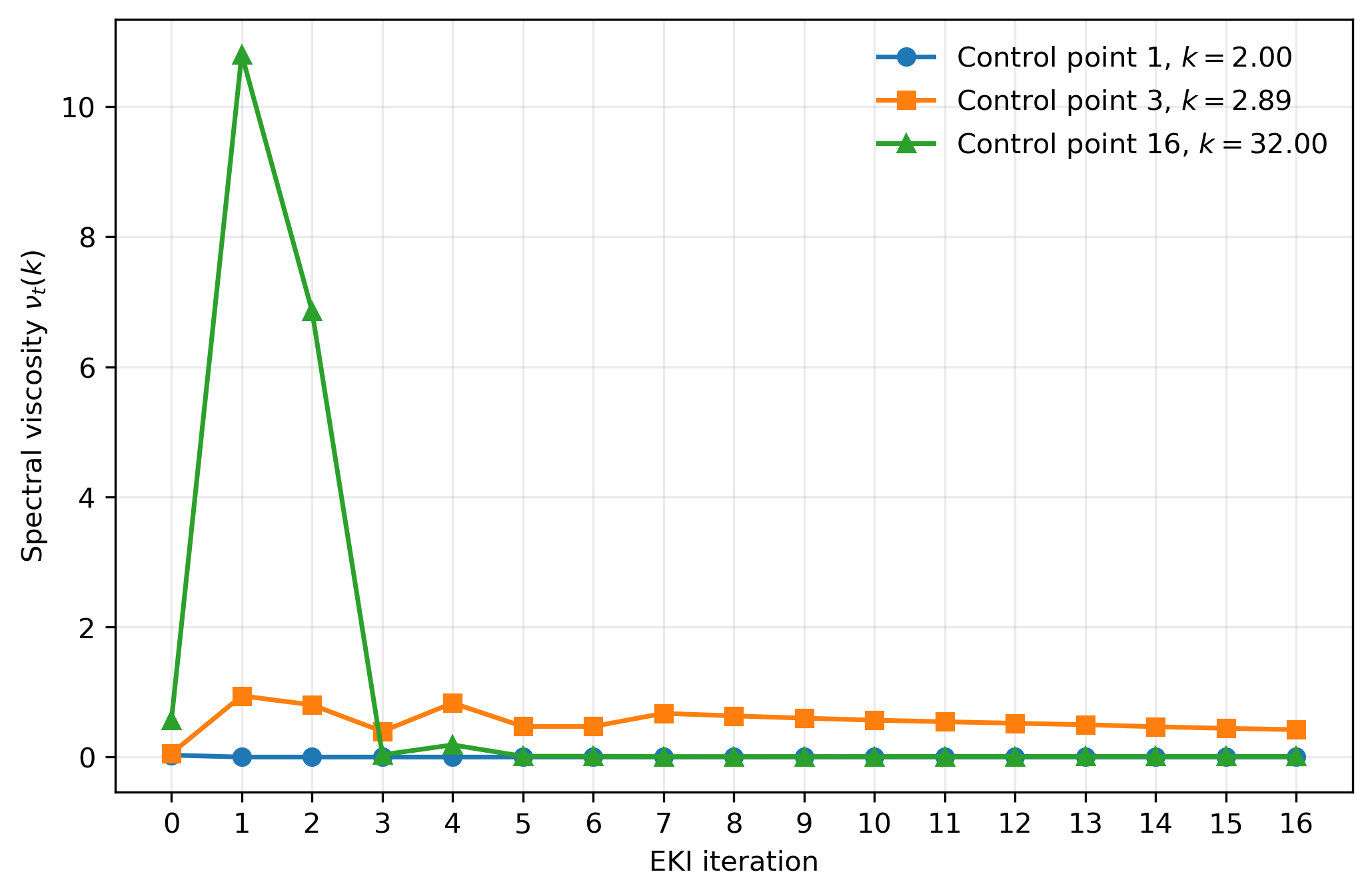}
        \caption{Selected control point histories.}
        \label{fig:control_history}
    \end{subfigure}

    \caption{EKI calibration history over 16 update iterations.
    (a) Evolution of the ensemble-mean spectral misfit between the LES ensemble and FDNS target. (b) Evolution of selected spectral viscosity control point values,
    displaying the large initial adjustment followed by progressive refinement of the learned closure.}
    \label{fig:eki_convergence}
\end{figure}

The trend of the discrepancy between the model and the empirical data, shown in Fig.~\ref{fig:eki_convergence}, can be used as a direct measure of how the EKI calibration evolves. The spectral misfit decreases rapidly during the initial iterations, indicating that the ensemble identifies parameter directions that improve agreement with the FDNS reference. Subsequent iterations produce smaller reductions as the ensemble approaches a region of parameter space associated with improved LES statistics.

The selected control point histories also illustrate that a large transient adjustment of an individual control point should not be interpreted as the final distribution of spectral dissipation. In particular, the high wavenumber control point undergoes a large early
adjustment before being progressively corrected during subsequent EKI iterations. The converged distribution of spectral viscosity is shown
separately in Fig.~\ref{fig:sv}.

The reduction of the misfit reveals that the covariance information inferred from the ensemble successfully establishes a meaningful relationship between the spectral viscosity parameters and the resulting LES observations. Importantly, the calibration achieves this reduction without requiring explicit derivatives of the LES solver with respect to the 16 closure parameters.
\subsection{Calibrated kinetic energy spectrum}

\begin{figure}[H]
\centering
{\includegraphics[width=0.68\textwidth]{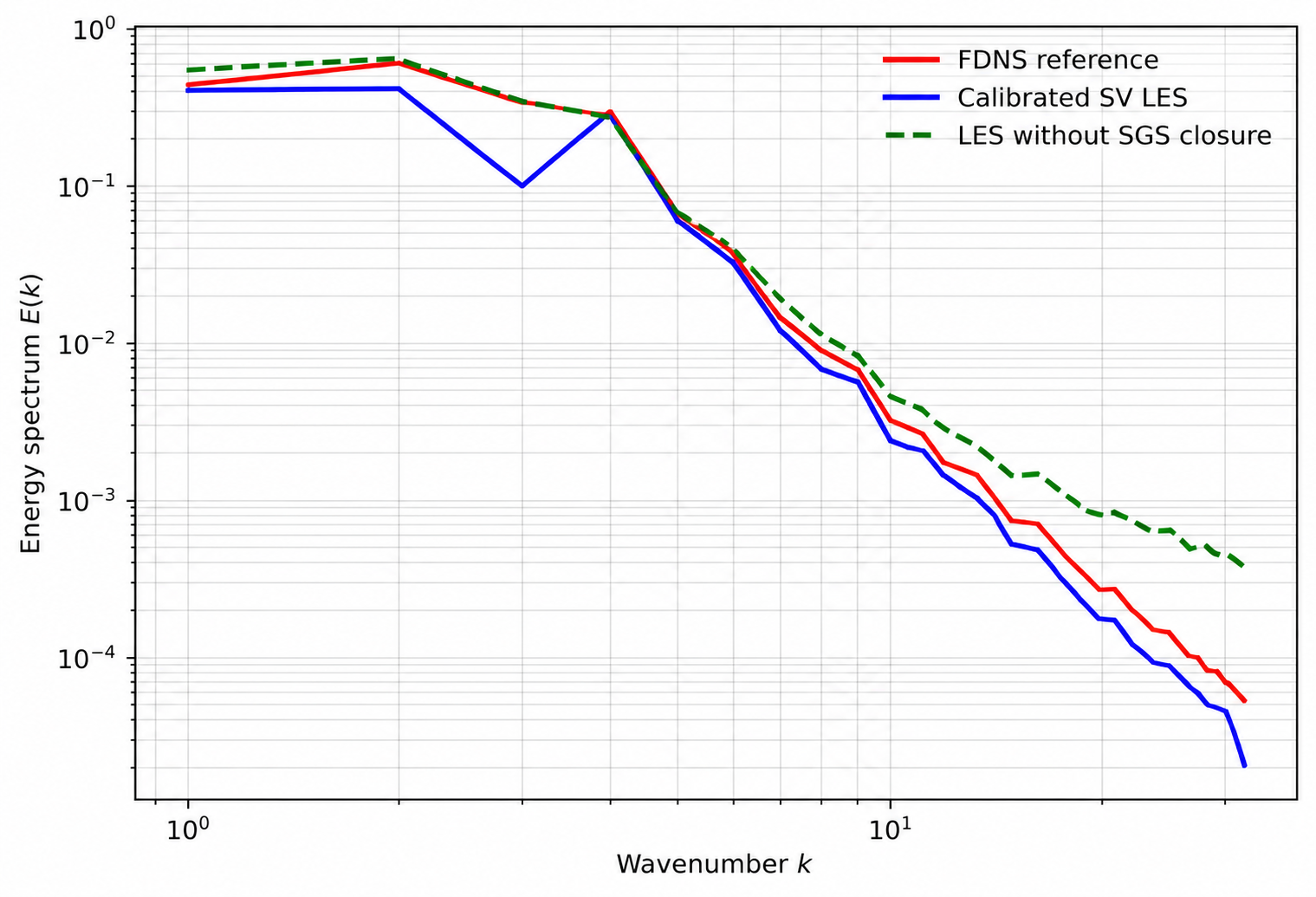}}
\caption{Time-averaged kinetic energy spectra for the FDNS reference, the LES with the calibrated spectral viscosity closure, and the LES without an SGS closure.}
\label{fig:final_spectrum}
\end{figure}

The kinetic energy spectrum produced by the calibrated spectral viscosity can be compared with the FDNS target as well as the LES performed without an SGS closure in Fig.~\ref{fig:final_spectrum} in order to assess the performance of the calibration on the LES statistics over the specified wavenumber range. The calculation of the LES without an SGS closure uses the same configuration but omits the spectral viscosity addition. It provides a baseline for evaluating the effect of the learned closure.

The LES without an SGS closure overpredicts the kinetic energy at higher resolved wavenumbers, which indicates insufficient dissipation of the smaller resolved scales. In comparison, the calibrated spectral viscosity substantially reduces this excess energy and more closely captures the overall spectral decay of the FDNS reference. Larger discrepancies occur at the lowest wavenumbers, where the LES can be seen to underestimate the energy compared to the reference solution. Nevertheless, the spectra show a good fit near the forcing region and follow the same trend of decay, especially at the high and intermediate wavenumbers. Over this range the calibrated LES spectrum remains predominantly below the FDNS target, which indicates that differences in the energy distribution still remain after calibration. 

Figs.~\ref{fig:eki_convergence} and \ref{fig:final_spectrum} show that EKI substantially reduces the spectral discrepancy relative to the no-SGS baseline. The learned closure suppresses the excess high-wavenumber energy and captures the overall spectral decay more accurately. However, its strong dissipation near the forcing wavenumber contributes to the remaining low-wavenumber energy deficit, indicating that further constraints on the spectral viscosity parameterization may be beneficial.

\section{Discussion}

The results demonstrate that EKI can be applied to an LES solver to estimate the parameters of a spectral viscosity closure by calibrating them against filtered high-resolution DNS target statistics. The calibration produces a substantial
reduction in spectral mismatch, and comparison with the no-SGS LES
demonstrates that the learned closure reduces excess energy at higher resolved wavenumbers. The calibrated LES captures the overall decay of the FDNS kinetic energy spectrum, although discrepancies remain at low
wavenumbers.

A key advantage of the approach is that the learned model remains physically interpretable, while also being derivative-free. This characteristic is particularly beneficial when the forward model is a nonlinear turbulent flow solver because it avoids the explicit computation of LES sensitivities with respect to the closure parameters. Instead, ensemble covariances provide an empirical relationship between closure parameters and the resulting LES kinetic energy spectrum.

Several limitations should nevertheless be recognized. First, the present investigation considers forced two-dimensional turbulence. Should the calibration framework be extended to three-dimensional turbulence, it would require a substantially higher computational expenditure. Second, the calibration relies primarily on the kinetic energy spectrum as the observational quantity. However, the discrepancies at the low wavenumbers suggest that a refinement of the calibration procedure or the spectral viscosity parameterization could be beneficial. Additional observations, alternative parameterizations, or regularization may therefore provide stronger constraints on the inferred viscosity profile. It is important to note, however, that since ensemble calibration requires multiple LES evaluations at every iteration, computational cost would increase with ensemble size and number of iterations. Although these coarse simulations are substantially less expensive than the corresponding DNS and can be evaluated independently, the total calibration cost would still increase considerably. Parallel evaluation of ensemble members and surrogate modeling are therefore reasonable possibilities for future computational optimization.

\section{Conclusions and perspectives}

This work investigated the online calibration of a parameterized spectral viscosity closure for LES of forced two-dimensional turbulence. A 16-dimensional parameter vector was chosen to shape the wavenumber-dependent viscosity profile, and EKI was employed to infer the closure parameters from a DNS-derived reference kinetic energy spectrum restricted by the wavenumber range resolved from the LES.

The EKI calibration substantially reduced the discrepancy between the LES and FDNS spectra and improved the resolved-scale energy distribution relative to the no-SGS baseline. These results support EKI as a derivative-free framework for calibrating interpretable turbulence closures from high-fidelity data.

Several further explorations naturally follow from the present work. The framework can be investigated at higher Reynolds numbers or extended to three-dimensional turbulence. Additionally, more adaptive spectral viscosity representations could be achieved by dynamically varying amplitudes or mechanisms permitting backscatter. A further perspective is the development of state-dependent closure strategies using reinforcement learning, which would provide a complementary approach in which closure behavior adapts dynamically to the evolving resolved flow.

\section*{Acknowledgements}

The authors gratefully acknowledge the support of the Department of Mechanical Engineering at Union College.

\end{document}